\documentclass[aps,jap,twocolumn,showpacs,longbibliography]{revtex4-1}
\pdfoutput=1
\usepackage{graphicx}
\usepackage{amsfonts}
\usepackage{amsmath}
\usepackage{natbib}

\begin{document}

\title{Rectification in mesoscopic AC-gated semiconductor devices}

\author{S.~P.~Giblin,$^{1}$ M. Kataoka,$^{1}$ J.~D.~Fletcher,$^{1}$ P.~See,$^{1}$ T.~J.~B.~M.~Janssen,$^{1}$
J.~P.~Griffiths,$^{2}$ G.~A.~C.~Jones,$^{2}$ I.~Farrer,$^{2}$ and D.~A.~Ritchie$^{2}$}

\affiliation{$^{1}$ National Physical Laboratory, Hampton Road, Teddington, Middlesex TW11 0LW, United Kingdom }

\affiliation{$^{2}$ Cavendish Laboratory, University of Cambridge, J J Thomson Avenue, Cambridge CB3 0HE, United Kingdom }
\email[stephen.giblin@npl.co.uk]{Your e-mail address}

\date{\today}

\begin{abstract}
We measure the rectified dc currents resulting when a 3-terminal semiconductor device with gate-dependent conductance is driven with an ac gate voltage. The rectified currents exhibit surprisingly complex behaviour as the dc source-drain bias voltage, the dc gate voltage and the amplitude of the ac gate voltage are varied. We obtain good agreement between our data and a model based on simple assumptions about the stray impedances on the sample chip, over a wide frequency range. This method is applicable to many types of experiment which involve ac gating of a non-linear device, and where an undesireable rectified contribution to the measured signal is present. Finally, we evaluate the small rectified currents flowing in tunable-barrier electron pumps operated in the pinched-off regime. These currents are at most $10^{-12}$ of the pumped current for a pump current of 100 pA. This result is encouraging for the development of tunable-barrier pumps as metrological current standards. 
\end{abstract}

\pacs{1234}

\maketitle
A broad class of experiments in condensed matter physics and nanoelectronics involve applying high-frequency ac signals to control gates on a mesoscopic semiconductor device. The control signals may be intended to drive charge pumping \cite{switkes1999adiabatic,blumenthal2007gigahertz,kaestner2008robust,fujiwara2008nanoampere}, induce photon-assisted transport \cite{kouwenhoven1994observation}, study the mesoscopic effects of microwave fields \cite{Liu1992Mesoscopic,dicarlo2003photocurrent,zhang2006directed}, or perform controlled operations on a quantum bit \cite{elzerman2004single}, to name just a few examples. If the experimental readout is in the form of a low frequency current or voltage, parasitic coupling between the ac-driven gate and other leads of the device can cause a rectified contribution to the measured signal \cite{brouwer2001rectification,dicarlo2003photocurrent,ferrari2005dc}. In some circumstances this contribution can mask the phenomenon under investigation \cite{switkes1999adiabatic,brouwer2001rectification}, or considerable effort may be required to isolate and subtract it \cite{dicarlo2003photocurrent,zhang2006directed}. A quantitative understanding of rectified current would clearly be a valuable experimental tool in this field. Several analyses of rectified current have been presented, \cite{dicarlo2003photocurrent,brouwer2001rectification,ferrari2005dc}, but none of them include all the relevent parameters, in particular the small dc bias which is inevitably present in a real experiment.

Here, we measure the rectified current in a very simple voltage-biased 3-terminal (source, drain and gate) device, a 2 micron-wide wire etched in a GaAs 2-Dimensional Electron Gas (2-DEG), crossed by a single metallic gate driven with an ac voltage. The rectified current displays a surprisingly rich behaviour as a function of the experimental control parameters dc bias voltage, dc gate voltage, and the amplitude of the ac gate voltage, and is particularly sensitive to the dc bias voltage. However, all of the observed behaviour is reproduced by a model based on the gate-dependent conductance, with a single fitting parameter characterising stray on-chip pick-up. 

As an application of our general method, we evaluate rectified contributions to the current in tunable-barrier (TB) electron pumps. These are 4-terminal devices, with source, drain, and two gates, one of which is driven with an ac signal at frequency $f$, resulting in a quantised dc current $I_{\text{P}}=ef$ \cite{kaestner2008robust,giblin2010accurate}. These pumps have demonstrated electron transfer accuracy at the 1 part-per-million level \cite{giblin2012towards}, and are possible candidates for a primary current standard following re-definition of the SI unit ampere \cite{milton2010quantum}. We find that the rectified current is less than a $10^{-12}$ correction to the pumped current, which is encouraging for further development of the TB pump as a current standard.

This paper is structured as follows: in section I we describe how rectified currents originate and present a method for modelling these currents in detail. In section II we describe our method for measuring the current. In section III we present measurements on a 3-terminal device, and show that the model developed in section I reproduces all of the observed features with a single fitting parameter. Finally in section IV we describe measurements on a TB pump which exbitits both rectified current and pumped current in different regions of parameter space, concluding that the rectified current at the optimal operating point for pumping is negligibly small.

\section{Model}

\begin{figure}
\includegraphics[width=8.5cm]{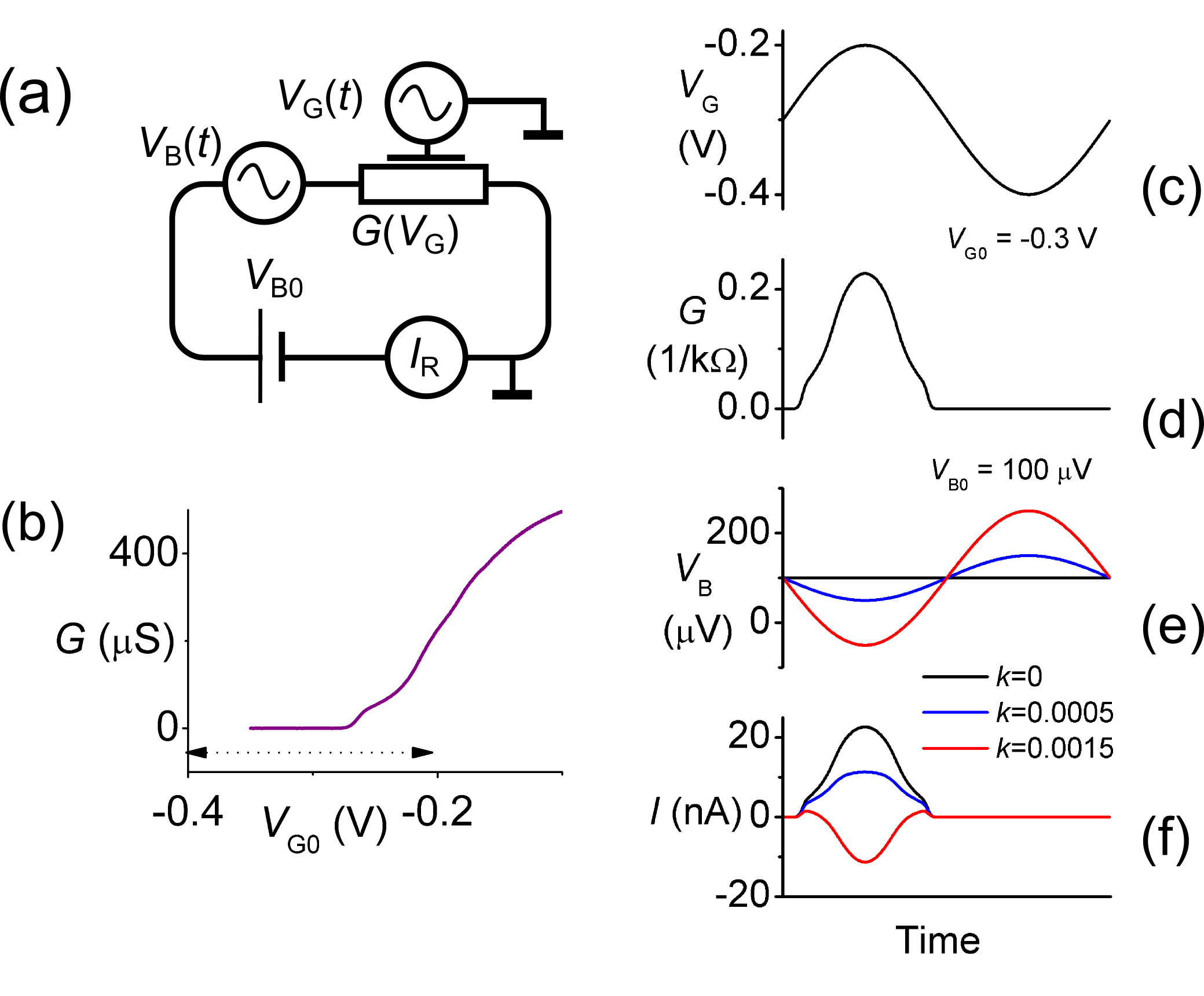}
\caption{\textsf{Rectification mechanism. (a) Schematic circuit elements used to model rectification. (b) Conductance as a function of $V_{\text{G}}$ with $V_{\text{AC}}=0$. The dotted line shows the range of gate voltage swept out with $V_{\text{G0}}=-0.3$~V, $V_{\text{AC}}=0.1$~V. (c-f) Instantaneous values of gate voltage, conductance, bias voltage and current, as a function of time for one cycle of the applied AC gate drive, calculated using (b) as input data. All four plots share the same time axis. Parameters are: $V_{\text{G0}}=-0.3$~V, $V_{\text{AC}}=0.1$~V, $V_{\text{B0}}=-100\thickspace\mu\text{V}$, $\phi=\pi$ and (g,h only) three values of the coupling constant $k$.}}
\label{fig:fig1}
\end{figure}

The equivalent circuit considered in this paper is shown in Fig. 1(a). We consider a device with a gate-voltage dependent conductance $G(V_{\text{G}})$ biased with a dc source-drain voltage $V_{\text{B0}}$. The gate is driven with a harmonic drive signal at frequency $f$ superimposed on a dc offset: $V_{\text{G}}(t) = V_{\text{G0}}+V_{\text{AC}}sin(2\pi ft)$, and we are interested in modelling the resulting dc current $I_{\text{R}}$ as a function of the experimental variables $V_{\text{G0}}$,$V_{\text{AC}}$,$V_{\text{B0}}$ and the DC conductance $G(V_{\text{G0}})$. Due to parasitic impedances, the gate couples to the source and drain leads, leading to an AC bias voltage across the device given by 

\begin{equation}V_{\text{B}}(t) = V_{\text{B0}}+kV_{\text{AC}}sin(2\pi ft+\phi)
\end{equation}

Where the amplitude and phase coupling constants $k$ and $\phi$ will generally be a function of frequency. Due to the large number of parasitic circuit elements present in a typical device chip, it will be difficult to determine $k$ and $\phi$ from first principles, so we will treat them as fitting parameters in our model. We include $V_{\text{B0}}$ in the analysis because current preamplifiers can have offset bias voltages as high as several hundred $\mu \rm V$ and in practice, even with offset nulling, it is difficult to obtain $V_{\text{B0}} \alt 10\mu \rm V$. 

In Fig. 1(b-f) we illustrate the mechanism for generating rectified current. The gate-dependent conductance [Fig. 1(b)] combined with the AC gate voltage (Fig. 1(c)) leads to a time-dependent conductance [Fig. 1(d)]. The presence of a time dependent bias voltage due to pickup [Fig 1(e)] results in a time-dependent current [Fig. 1(f)] and a dc rectified current given by the average of this current over one cycle of the applied gate voltage \cite{brouwer2001rectification,dicarlo2003photocurrent}: 

\begin{equation} I_{\text{R}}=f\int_{0}^{1/f} V_{\text{B}}(t)G(t)dt
\end{equation}

Note that in the absense of stray coupling ($k=0$), a rectified current is still present due to the non-zero $V_{\text{B0}}$ [black line in Fig. 1(f)]. As the stray coupling is progressively turned on, for the parameters illustrated in the figure $I_{\text{R}}$ changes sign. Generally with non-zero $V_{\text{B0}}$ and $k$, the rectified current can change sign as any of the experimental parameters are varied. Furthermore, for small $V_{\text{AC}}$, $I_{\text{R}}(V_{\text{G0}})\propto \partial^{2} G/\partial V_{\text{G}}^{2}|_{V_{\text{G}}=V_{\text{G0}}}$. One might naiively expect $I_{\text{R}} \propto \partial G/\partial V_{\text{G}}$ because a steeper slope to $G(V_{\text{G0}})$ results in a larger excursion in $G(t)$ for a given $V_{\text{AC}}$. However, referring to Fig. 1, and considering the case of a linear slope to $G(V_{\text{G0}})$, $G(t)$ becomes a triangle wave, and the two half-cycles of $V_{\text{B}}(t)$ give rise to equal and opposite contributions to $I(t)$, which sum to zero leaving the trivial contribution, $I_{\text{R}}=V_{\text{B0}}G|_{V_{\text{G}}=V_{\text{G0}}}$.

Substituting Eq.(1) into Eq.(2), we obtain

\begin{align}
I_{\text{R}}&=f \biggl[\int_{0}^{1/f} V_{\text{B0}}G(t)dt \nonumber\\*
&+ V_{\text{AC}}kcos(\phi)\int_{0}^{1/f} G(t)sin(2\pi ft)dt \biggr]
\end{align}

In deriving Eq.(3) we have made use of standard trigonometric identities, and the fact that $\int G(t)cos(2\pi ft)dt = 0$ over one cycle. From Eq.(3), it is clear that measurement of $I_{\text{R}}$ cannot determine $k$ and $\phi$ independently, as they appear in the equation in the form of the product $kcos\phi$. Each set of $[k,\phi]$ satisfying $kcos\phi$=constant is associated with a different $I(t)$ waveform (Fig. 1(f)), but with the same average current. However, this limitation will not have a serious impact on the overall conclusions of this study. Equations (1) and (2), combined with a measurement of $G(V_{\text{G}})$, should now allow us to model the rectified current in a real device, treating the term $kcos\phi$ as a fitting parameter. In the next section we describe our device, and the method for measuring $I_{\text{R}}$.

\section{Experimental Methods}

A schematic illustration of our 3-terminal device is shown in Fig 2(a). A $2$~$\mu$m wide conducting wire is etched in a GaAs 2-Dimensional Electron Gas (2-DEG), and crossed by a metallic gate. Applying a negative voltage to this gate depletes the 2-DEG beneath, suppressing conductance through the channel. In Fig. 2(b) we show an optical microscope image of the sample chip. The gate is the termination of the vertical finger running from the top of the picture. The finger coming from the bottom edge, and four additional structures pointing from the corners of the chip towards the centre are not used in this device. The stray capacitances from the gate to the left (source) and right (drain) leads, $C_{\text{GS}}$ and $C_{\text{GD}}$, are indicated, along with the impedances $Z_{\text{S0}}$ and $Z_{\text{D0}}$ from the source and drain leads to ground. When an ac voltage is applied to the gate, the stray impedances will operate as potential dividers and cause parasitic ac voltages to appear on the source and drain terminals. The impedances $Z_{\text{S0}},Z_{\text{D0}}$ are formed from a combination of 2-DEG resistance, capacitance to the ground plane, resistance and inductance of optically-defined metal tracks, bond wire inductance, and other parasitic components which are difficult to model quantitatively. A full model of the equivalent circuit of the chip is beyond the scope of this study. However, based on the highly symmetric chip layout, we can assume $(C_{\text{GS}} \approx C_{\text{GD}})$, $(Z_{\text{S0}} \approx Z_{\text{D0}})$. Furthermore, one plausible model for $Z_{\text{S0}},Z_{\text{D0}}$ is that they are mainly capacitive, being formed by the capacitances between the 2-DEG and surrounding grounded metal structures. In this case, the voltages induced on the source and drain leads will be in phase with the gate voltage, and independent of frequency, yielding $k\ll 1, \phi\in\left\{0,\pi\right\}$. In other words, the voltage of both the source and drain leads will oscillate with roughly the same amplitude and phase with respect to $V_{\text{AC}}$. The voltage \textit{difference} between source and drain will be much smaller than $V_{\text{AC}}$ and either in-phase or out of phase, depending on whether the larger pick-up is on the source or drain side.

\begin{figure}
\includegraphics[width=6.5cm]{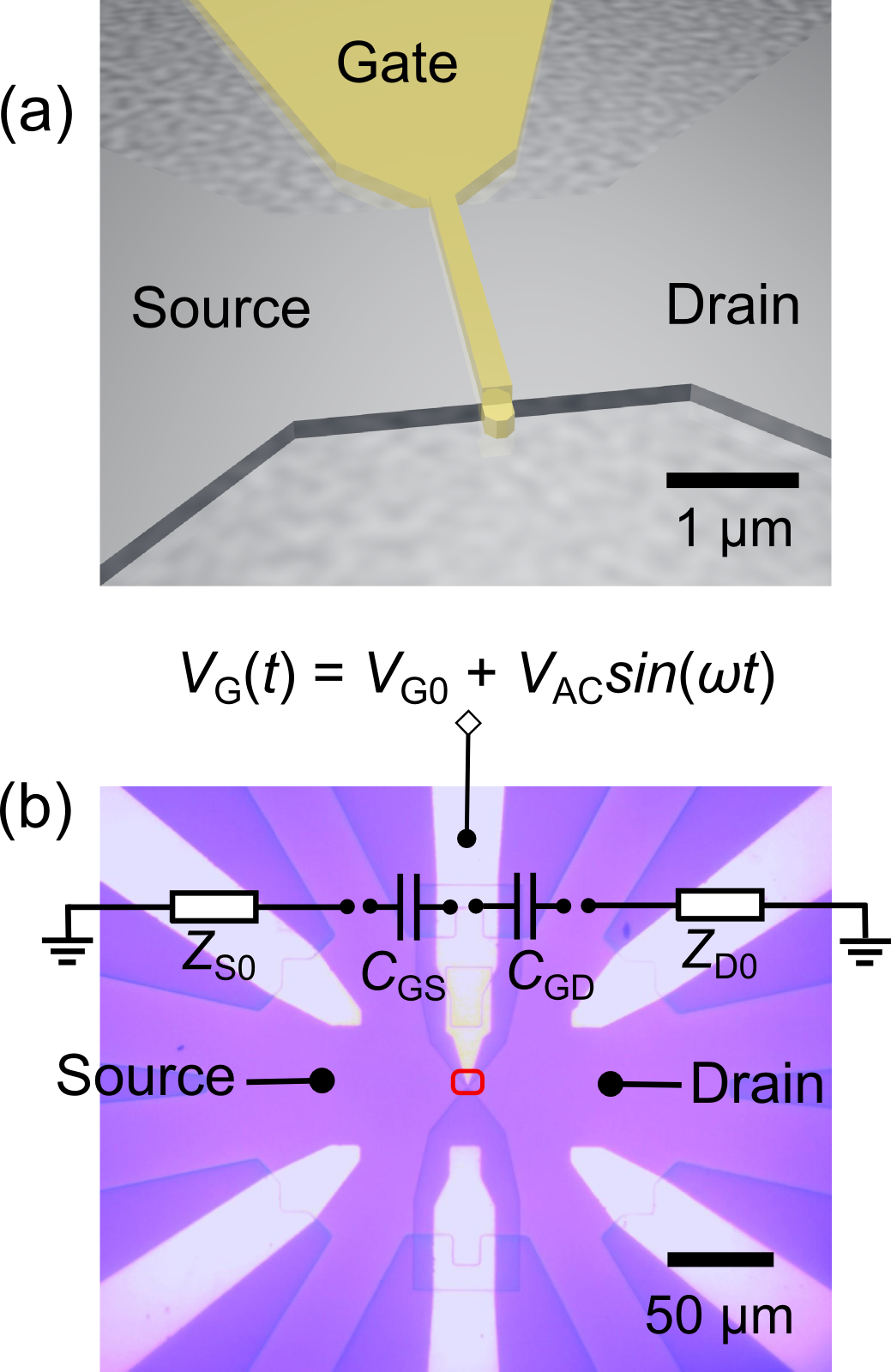}
\caption{\textsf{Device details. (a): Illustration of the 3-terminal test device, showing the etched channel running from left to right, crossed by a metallic gate. (b) Optical microscope image of the device chip. The image in (a) is a zoom of the central region (red box). The metallic gates show up as white, and the region where the 2-DEG has been removed by etching is dark purple. Stray impedances between the gate, source, drain and ground are indicated.}}
\label{fig:fig2}
\end{figure}

We fabricated our devices on GaAs/Al$_{x}$Ga$_{1-x}$As wafers using standard techniques: wet-etching to define the channel, and electron-beam lithography for the metallic gates \cite{blumenthal2007gigahertz}. The samples were cooled to a temperature of $4.2$~K by immersing in liquid helium and the current was measured using a lock-in technique similar to one described previously \cite{dicarlo2003photocurrent}. The AC gate drive came from an RF source with output power $P_{\text{RF}}$ calibrated in dBm for a $50\thickspace\Omega$ load. The output of the RF source was supplied to the gate through a room-temperature bias-tee and a $1$~m length of stainless-steel semi-rigid co-axial cable. The RF source was chopped with a square wave, duty cycle $0.5$, at a frequency $f_{\text{L}}\approx 300$~Hz. The source lead of the device was connected to a voltage bias source, and the drain lead was connected to an ac-coupled transimpedance amplifier. The ouptut of this amplifier was monitored by a lock-in amplifier referenced to $f_{\text{L}}$. The lock-in does not measure $I_{\text{R}}$ directly, but a current $I_{\text{M}}$ which is proportional to the difference between $I_{\text{R}}$ and the current flowing when $V_{\text{AC}}=0$:

\begin{equation} I_{\text{M}}=\frac{\sqrt{2}}{\pi}\thickspace\Bigl[I_{\text{R}}-V_{\text{B0}}G_{\text{0}}\Bigr]
\end{equation}

Where $G_{\text{0}}$ is the conductance at $V_{\text{G}}=V_{\text{G0}}$. The pre-factor arises because the input to the lock-in amplifier is a square wave, and the lock-in measures the RMS amplitude of the first harmonic of this square wave. The conductance $G(V_{\text{G0}})$ was measured in a separate experiment using a standard lock-in technique with a small low-frequency AC bias voltage. The true bias voltage $V_{\text{B0}}$ across the device was determined by measuring the DC current as a function of applied bias, with $V_{\text{G0}}=0$~V. This measurement was repeated several times each day to correct for drifts in thermo-electric potentials and the amplifier offset.

\begin{figure}
\includegraphics[width=9cm]{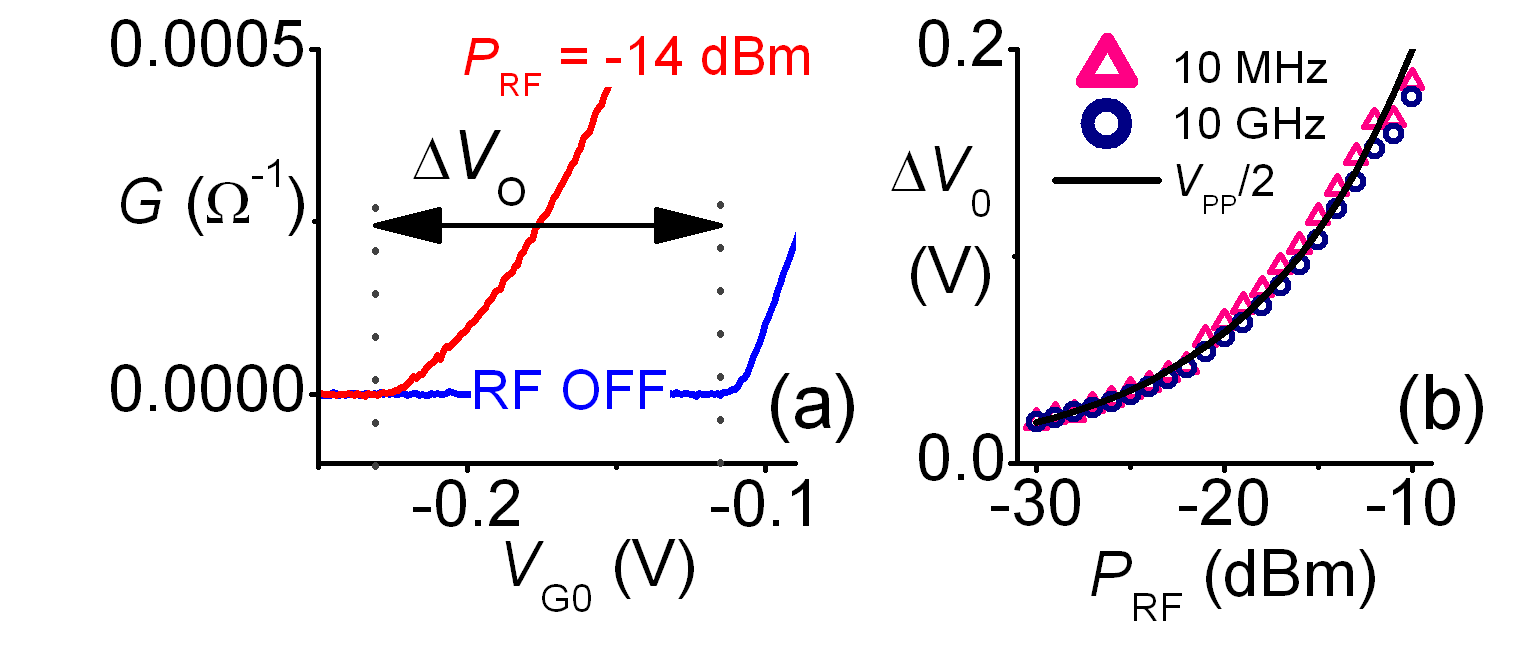}
\caption{\textsf{Calibration of AC voltage at the device gate. (a): Conductance as a function of DC gate voltage with $V_{\text AC}$ off (blue line) and -14 dBm RF power applied from the source at $f=10$~GHz (red line). The shift in conductance onset $\Delta V_{\text{0}}$ due to applying the RF drive  is indicated. (b): points: values of $\Delta V_{\text{0}}$ obtained from data similar to that shown in (a) over a range of RF power levels, at $f=10$~MHz (triangles) and $10$~GHz (circles). For comparison, the solid line shows the expected voltage at the sample gate calculated from the output power of the RF source and assuming a perfect open circuit termination. Small discontinuities in $\Delta V_{\text{0}}$ at $P_{\text{RF}}=-12$ and $-22$~dBm result from range switching attenuators internal to the RF source.}}
\label{fig:fig3}
\end{figure}

In our experiment, the device gate presents approximately an open circuit termination to the RF transmission line. In the ideal case of an open circuit, we expect half the peak-peak gate voltage to be given by $V_{\rm{PP}}/2 = 2\sqrt{2}\medspace\sqrt{1\medspace\text{mW}\times 50\medspace\Omega}\medspace\times10^{P_{\text{RF}}/20}$. However, because of the possibility of multiple reflections, we calibrated $V_{\text{AC}}$ in terms of the generator output power $P_{\text{RF}}$ using the shift in the conductance pinch-off voltage as a direct measure of $V_{\text{AC}}$. In Fig. 3(a) we show the conductance as a function of gate voltage measured with the RF source turned off (blue line) and with $f=200$~MHz, $P_{\text{RF}}=-14$~dBm (red line). The shift in conductance pinch-off is denoted $\Delta V_{\text{0}}$. In Fig. 3(b), we plot $\Delta V_{\text{0}}(P_{\text{RF}})$ for $f=10$~MHz and $f=10$~GHz (symbols) along with $V_{\text{PP}}/2$ (solid line). Overall there is very good agreement between $\Delta V_{\text{0}}$ and $V_{\text{PP}}/2$ over a wide frequency range, which shows that our cryogenic transmission line is free of multiple reflections. A very small reduction in $\Delta V_{\text{0}}$ at higher frequency is visible, due to the attenuation of the co-axial line. In section IV, where we employ frequencies up to 1.4 GHz, we introduce an attenuation factor in order to fit $I_{\text{M}}$ to our model. In section III the frequency was $100$~MHz, and it was sufficient to set $V_{\text{AC}}=V_{\text{PP}}/2$.



\section{Experimental Data}

We measured $I_{\text{M}}$ for the device illustrated in Fig. 2 over a wide range of parameters $[f,V_{\text{B0}},V_{\text{G0}},P_{\text{RF}}]$. We found that $I_{\text{M}}$ was independent of $f$ over the entire experimental range $10$~kHz$\leq f\leq 10$~GHz, apart from changes in amplitude attributed to attenuation in the RF transmission line. This suggests that the simplified model for the equivalent circuit proposed in section II is a good approximation for this sample. In Fig. 4(a) (symbols) we present data, at $f=100$~MHz, of $I_{\text{M}}(V_{\text{G0}})$ close to the conductance pinch-off, for three values each of $V_{\text{B0}}$ and $P_{\text{RF}}$. As $P_{\text{RF}}$ is increased, the onset of rectified current is shifted towards more negative gate voltages in the same way as the conductance onset illustrated in Fig. 3(a). The data at zero bias voltage shows clearly that a finite bias is not required to drive a current, due to the asymmetric stray coupling from the gate to the source and drain leads breaking the symmetry of the device. The solid lines show the best fit to equations (1,2,4) with $k=0.00038$ and $\phi=\pi$ ($\phi=0$ did not yield good fits), using the measured $G(V_{\text{G0}})$ shown in Fig. 1(b) as input data. The fit was performed by simultaneously minimising $\chi^2$ for all 9 $I_{\rm {M}}(V_{\text{G0}})$ data sets shown in the figure. Small discrepancies between the data and the fits could be removed by allowing $V_{\text{B0}}$ to be an extra fit parameter for each $I_{\rm {M}}(V_{\text{G0}})$ data set. This reflects the real experimental situation in which $V_{\text{B0}}$ could drift by a few $\mu \rm V$ over the time-scale required to acquire data. The value of $k$ is consistent with the high level of symmetry in the device, which implies $k\ll 1$, and is similar to the value measured in another experiment \cite{dicarlo2003photocurrent}. 

\begin{figure}
\includegraphics[width=8.5cm]{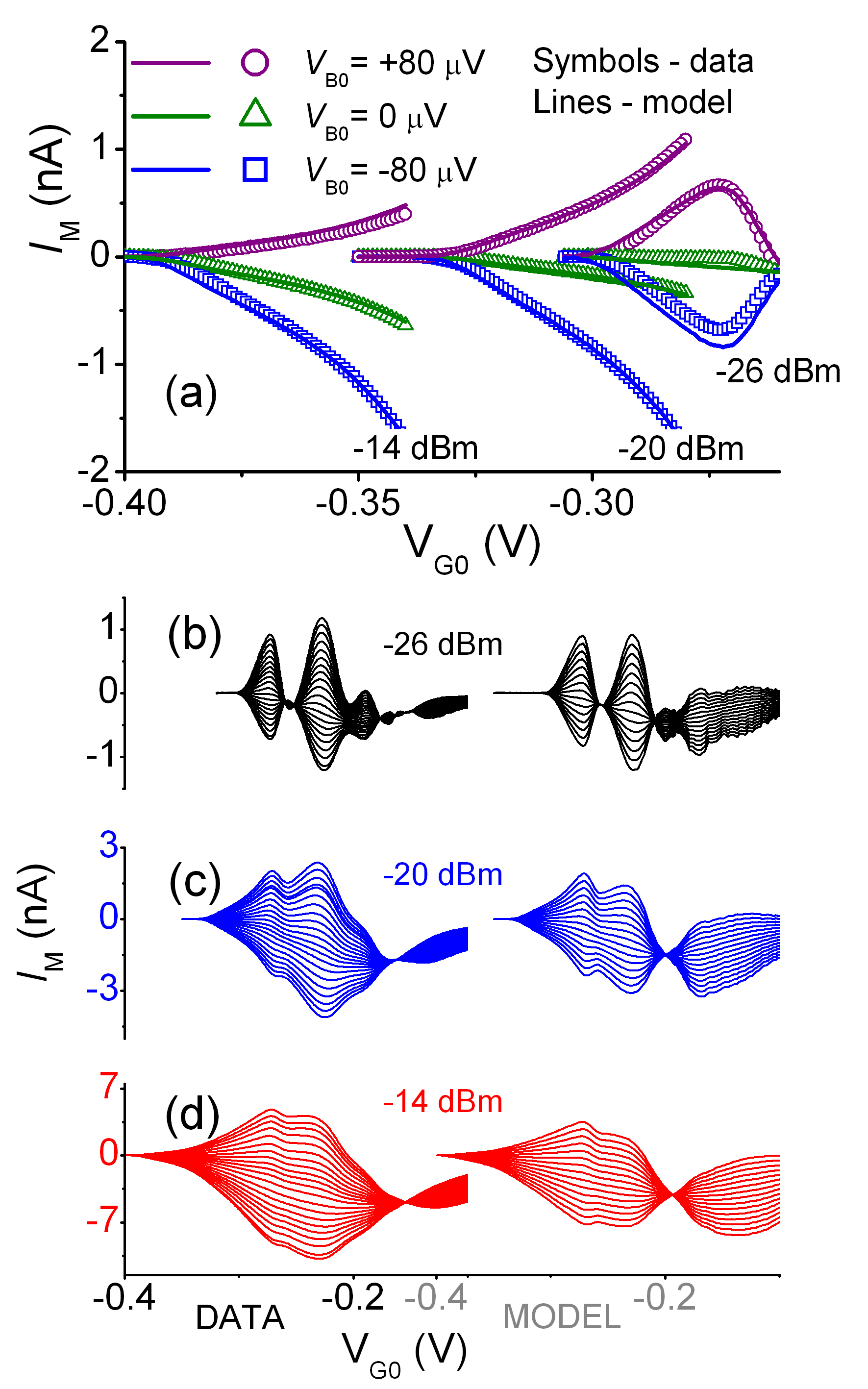}
\caption{\textsf{Data and fits for the device illustrated in Fig. 2. (a) current (symbols) measured for three values each of $V_{\text B0}$ and applied RF power at $f=100$~MHz. The lines show fits to equations (1,2,4). (b-d): Comparison between $I_{\text M}(V_{\text G0})$ data and model fit similar to (c) but over a wider range of $V_{\text G0}$. Left panels show experimental data and right panels show model calculations. In each panel, $V_{\rm B0}$ is stepped from $-0.1$ to $0.1$~mV in increments of $10$~$\mu$V (data) and $12.5$~$\mu$V (model).}}
\label{fig:fig3}
\end{figure}

\begin{figure*}
\includegraphics[width=17cm]{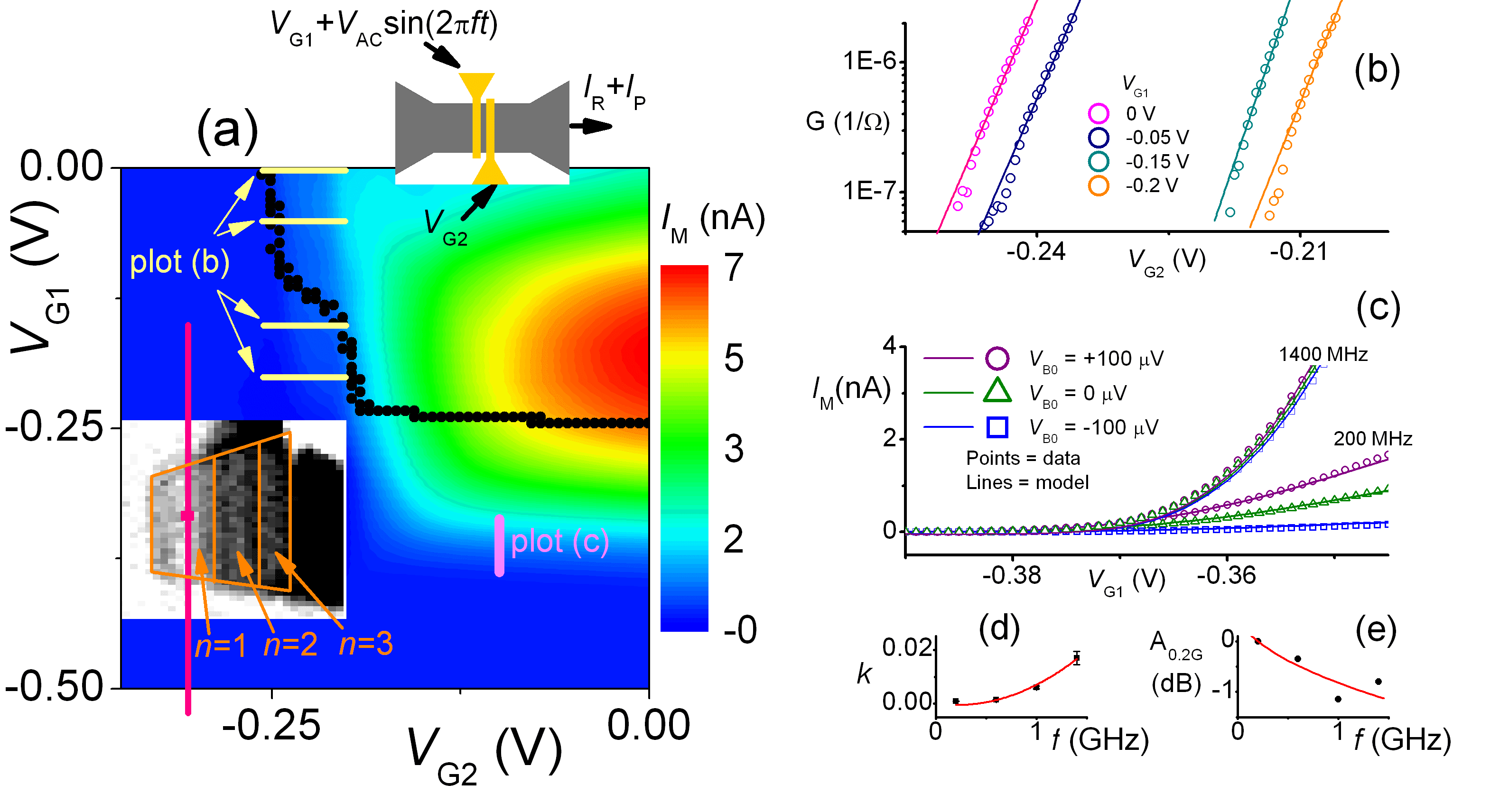}
\caption{\textsf{Rectifcation in a two-gate TB pump. (a): Total current as a function of both DC gate voltages, with $P_{\text{RF}}= -11$~dBm and $f=200$~MHz. The data set is plotted as a colour map apart from the region $-0.42\thickspace\text{V}\leq V_{\text{G1}}\leq -0.24\thickspace\text{V}$, $-0.35\thickspace\text{V}\leq V_{\text{G2}}\leq -0.20\thickspace\text{V}$ which is plotted as a grey-scale derivitive $dI_{\text{M}}/dV_{\text{G2}}$ (white = zero) to highlight the quantised pump current. The quantised plateaus with $n=1-3$ are demarcated with orange lines for clarity. The pink square shows the operating point of the pump in the middle the $n=1$  plateau, and the vertical pink line intersecting this point shows the expected range of parameter space traced out when the entrance gate is driven with $P_{\text{RF}}= -11$~dBm. Black points: onset of conductance with the RF source turned off. $G>10^{-6}\mu\text{S}$ above and to the right of this onset line. Horizontal yellow and short vertical pink lines show the range of plots (b) and (c) respectively. The inset is a schematic illustration of the pump (grey = 2-DEG), yellow = metallic gates). (b): Conductance as a function of $V_{\text{G2}}$ at several values of $V_{\text{G1}}$. Solid lines are fits to an exponential function. (c) Current as a function of $V_{\text{G1}}$ at 3 values of DC bias voltage and 2 drive frequencies. Points are measured data and lines are fits to equations (1) and (2). (d,e): Fit parameters $k$ and $A_{\text{0.2G}}=A(f)-A\text{(0.2 GHz)}$ obtained from the data in (c) and additional data at $f=600$~MHz and $f=1$~GHz. The solid line in (d) is a fit to a quadratic function , and the line in (e) shows the attenuation of the transmission line from manufacturer's data.}}
\label{fig:fig4}
\end{figure*}

In Fig 4(b-d), left panels, we show data over a wider range of gate voltage, which shows more clearly the non-trivial dependence of the rectified current on the experimental parameters. The structure in $I_{\text{M}}(V_{\text{G0}})$ derives entirely from fine features in $G(V_{\text{G0}})$. When $V_{\text{AC}}$ is less than the characteristic scale of features in $G(V_{\text{G0}})$, as is the case for $P_{\text{RF}}=-26$~dBm, $I_{\text{R}}$ is proportional to $d^{2}G/dV_{\text{G0}}^{2}$ as expected. 
The right panels show the current calculated from Eqs. (1,2,4) using the same $[k,\phi]$ fitting parameters extracted from the data of Fig. 4(a). We stress that the fit is constrained only by the current close to the pinch-off gate voltage, and it still captures very well the main features of the data. The model loses accuracy at more positive gate voltages, where the conductance of the device is large. This is because $G(V_{\text{G0}})$ was measured using a two-terminal method which includes a contribution from the ohmic contacts to the 2-DEG, so the model will tend to over-estimate $I_{\text{R}}$ for more postive $V_{\text{G0}}$ where the contacts become a significant contribution to the overall 2-terminal resistance. 

We have shown that rectified currents can be accurately modelled in a simple 3-terminal device, using the measured conductance characteristic and a single fitting parameter characterising the strength of the stray pickup between the gate and source, drain terminals. In the next section we consider a more experimentally relevent case, where rectified current is present alongside current due to another mechanism, in this case quantised charge pumping, and we wish to distinguish the two contributions.

\section{The two-gate tunable-barrier pump}

The single-parameter tunable barrier pump \cite{kaestner2008robust} is illustrated schematically in the upper inset to Fig 5(a). It differs from the device discussed in the previous section by having a second gate crossing the channel. The gates are biased with negative DC voltages $V_{\text{G1}}$ and  $V_{\text{G2}}$, and one of the gates is additionally driven with an AC voltage. This has the effect of pumping electrons over the potential barrier created by the fixed gate without the need for a source-drain bias voltage \cite{kaestner2008robust,kaestner2008single,giblin2010accurate}. The pumped current $I_{\text{P}}$, and the rectified current $I_{\text{R}}$ will be present in different proportions for different values of the fixed gate voltages, and we are mainly interested in evaluating $I_{\text{R}}$ at the optimal pump operating point, where we wish for it to be a negligible (less than $1$ part in $10^{8}$) correction to the pump current. One motivation for this investigation was the fact that in some studies of the metrological accuracy of pumps \cite{giblin2010accurate,giblin2012towards}, relatively large values of $V_{\text{AC}}$, up to $0.5$~V, were required to attain the optimal pumping regime. This implies that the pump state traverses a large region of the gate voltage plane during the operation cycle, and is in contrast to the operation of normal state metallic \cite{keller1996accuracy}, and hybrid normal/superconducting pumps \cite{pekola2007hybrid} in which the AC gate voltages are in the mV range. 
We investigated the pump using the methods described in sections I and II, bearing in mind that the conductance $G(t)$ in Eqn. (2) is now a function of both gate voltages. Fig. 5(a) shows the measured current as a colour map in the $[V_{\text{G1}},V_{\text{G2}}]$ plane, with$P_{\text{RF}}=-11$~dBm and $f=200$~MHz. Additionally, the onset of conductance with $V_{\text{AC}}=0$ is shown on the same axes, as black points separating $G>2\mu$S (upper and right) and $G<2 \mu$S (lower and left) regions. The plot of $I_{\text{M}}(V_{\text{G1}},V_{\text{G2}})$ is dominated by a large rectified signal in the region where the device conductance is appreciable. The pump current $I_{\text{P}}\approx 32$~pA for the $n=1$ plateau is indistinguishable from zero on the scale of the main colour map, so the part of the data set where quantised pumping occurs has been plotted as a greyscale derivitive map. The first 3 quantised plateaus are visible in the derivitive plot as lighter-coloured regions (highlighted with orange lines for clarity), and a pink square marks the optimal operating point for quantised pumping on the $n=1$ plateau. 

With reference to this figure, the problem of rectified currents can be clearly stated. The operating point of the pump $(V_{\text{G1}},V_{\text{G2}})=(-0.33,-0.32)$~V is far into the pinched-off (Small $G$) region of the diagram, but the AC gate drive added to $V_{\text{G1}}$ takes the point $p$ representing the instantaneous state of the pump (long pink line) closer to the conductance edge. To calculate the rectified current resulting from the excursion of $p$ in gate voltage space, we need to know the value of $G$ along that line, and also the size of the AC bias voltage across the pump. For the latter, we will use the methodology described in the previous sections of the paper. 

First, we consider the conductance $G$. In Fig. 5(b) we show measurements of $G(V_{\text{G2}}$) for various values of $V_{\text{G1}}$. The data have been fitted to an exponential function, $G$ dropping by a factor of $10$ for a $6$~mV reduction in $V_{\text{G2}}$. The exponential dependence is expected for transport dominated by tunneling through a potential barrier with height proportional to the magnitude of $V_{\text{G2}}$. The tunnel conductance is expected to fall off slightly more rapidly than a pure exponential because making $V_{\text{G2}}$ more negative causes a widening as well as a raising of the barrier. This effect is just visible in the data of Fig. 3(b). A consequence is that our extrapolation of $G$ to lower values will tend to be an over-estimate.

Next, we measured $I_{\text{M}}(V_{\text{G1}})$ for $V_{\text{G2}}$ sufficiently positive that the barrier formed by gate 2 is too low for quantised pumping to occur. Examples of such data, at $V_{\text{G2}}=-0.1$~V, are shown in Fig. 5(c) (points) for 3 values of $V_{\text{B0}}$ and 2 values of $f$. In contrast to the single-gate device studied in the previous section, here the current is frequency dependent. As the frequency is increased, the overall magnitude of the current increases, and changing $V_{\text{B0}}$ has less effect. However, the onset value of $V_{\text{G1}}$ where the current begins to rise from zero is only weakly frequency dependent. We fitted the data in Fig. 5(c) to Eqs. (1,2,4) using $G(V_{\text{G1}})|_{V_{\text{G2}}=-0.1~V}$ measurements as input data. For each frequency, we fitted data at the different values of $V_{\text{B0}}$ as a single data set, setting $\phi=0$, and minimising $\chi^2$ in the $[k,A]$ plane, where $A$ is the the attenuation of the microwave transmission line in dBm: $V_{\text{AC}}=V_{\text{P/2}}(P_{\text{RF}}-A)$. We obtained $k=0.00088\medspace(f=200\thickspace\text{MHz})$ and $k=0.017\medspace(f=1.4\thickspace\text{GHz})$. Values of $k$ and $A$ extracted from the fits at 4 frequencies are plotted in Figs. 5(d) and 5(e). The fitted attenuation increases with frequency as expected for 1 m of semi-rigid stainless-steel co-axial cable [solid line in Fig. 5(e)].

The coupling parameter $k$ [Fig. 5(d)] increases quadratically with frequency. A full quantitative comparison of this behaviour with a circuit model of the device was not possible because not all on-chip stray impedances were measurable in isolation. However, we obtained qualitative agreement in the frequency range $100\thickspace\text{MHz}\leq f\leq 2\thickspace\text{GHz}$ using measured values of stray capacitances $C_{\text{GS}}=40$~fF, $C_{\text{GD}}=60$~fF and a plausible circuit model for $Z_{\text{S0}}$ and $Z_{\text{D0}}$ implemented in a SPICE simulation. Our model for $Z$ consisted of a series connection of four components: a 2-DEG resistance $\sim100\thickspace\Omega$, ohmic contacts $1\thickspace\text{k}\Omega$ in parallel with $10$~pF, bond wires $\sim1$~nH and cable capacitance $\sim100$~pF. Allowing equivalent parameters on source and drain sides of the device to differ randomly by up to $30\thickspace\%$ yielded a frequency-dependent $k$ and roughly frequency-independent $\phi$ in good agreement with the data of Fig. 5(d). 

The two samples investigated in this study, the single-gate sample and the two-gate pump, were fabricated using the same optically-defined mask. The different frequency-dependent behaviour seen in the two samples is then most likely due to the different bond wire inductances. The length and placement of bond wires was an uncontrolled parameter, with the length varying by up to a factor of $4$, depending on the exact placement of the sample chip within the holder. This hypothesis could be confirmed by experiments in which the bond wire length was controlled.
We also considered the effect of a frequency-dependent cross-coupling between the two gates, which would distort the trajectory of $p$ from a line at constant $V_{\text{G2}}$ to an ellipse in the $V_{\text{G1}},V_{\text{G2}}$ plane. This hypothesis was discarded, at least in the frequency range of the data in Fig. 5, $200$~MHz$\leq f\leq 1400$~MHz, because the best fit to the data of Fig. 5(c) was always obtained with $p$ moving with constant $V_{\text{G2}}$.
Finally, with $G(V_{\text{G2}})$ derived from the fits in Fig. 5(b), and $k(f)$ from Fig. 5(d), we used equations (1) and (2) to calculate $I_{\text{R}}$  for the pump at the operating point. Unsurprisingly, for $P_{\text{RF}}=-11$~dBm, this current is vanishingly small, of order $10^{-30}$~A. This is a direct consequence of the steep drop in conductance as $V_{\text{G2}}$ is made more negative. However, in this particular sample, the conductance pinch-off shifts to more negative values of $V_{\text{G2}}$, as $V_{\text{G1}}$ is made more positive. This means that a fairly modest increase in $P_{\text{RF}}$ would bring the point $p$ much closer to the conductance edge. Using $P_{\text{RF}}=-6$~dBm and $k=0.017$ gives $I_{\text{R}}\sim 10^{-22}$~A, or 1 part in $10^{12}$ of a pump current of $100$~pA.

To conclude, we have systematically studied the rectified current in semiconductor devices with a gate-voltage-dependent conductance, which appears when the gate is driven with an ac signal. We have shown quantitative agreement with a simple model based on the measured conductance of the device, and a fitting parameter which describes the strength of the stray coupling between the gate and the source, and drain leads. This is a valuable tool in experiments employing ac gate drive, where rectified current may constitute an artefact signal. More work is required to understand the frequency dependence of the rectified current, in particular the role of asymmetries in the chip layout and bond wires. 


\begin{acknowledgments}
This research was supported by the UK department for Business, Innovation and Skills and the European Metrology Research Programme, grant no. 217257. SPG thanks M. Connolly for a careful reading of the manuscript.
\end{acknowledgments}

\bibliography{SPGrefs_RectPaper}

\end{document}